\documentclass[prb,twocolumn,superscriptaddress,epsfig]{revtex4}
\usepackage{graphicx}
\usepackage{dcolumn}
\usepackage{bm}
\begin{document}
\title {Transport Through an Electrostatically Defined Quantum Dot Lattice \\in a Two-Dimensional Electron Gas}
\author{Srijit~Goswami}
\email{goswami@physics.iisc.ernet.in}
\affiliation{Department of Physics, Indian Institute of Science, Bangalore 560 012, India.}
\author{M.~A.~Aamir}
\affiliation{Department of Physics, Indian Institute of Science, Bangalore 560 012, India.}
\author{Christoph~Siegert}
\affiliation{Cavendish Laboratory, University of Cambridge, J.J. Thomson Avenue, Cambridge CB3 0HE, United Kingdom.}
\author{Michael~Pepper}
\affiliation{Department of Electronic and Electrical Engineering, University College, London WC1E 7JE, United Kingdom}
\author{Ian~Farrer}
\affiliation{Cavendish Laboratory, University of Cambridge, J.J. Thomson Avenue, Cambridge CB3 0HE, United Kingdom.}
\author{David~A.~Ritchie}
\affiliation{Cavendish Laboratory, University of Cambridge, J.J. Thomson Avenue, Cambridge CB3 0HE, United Kingdom.}
\author{Arindam~Ghosh}
\affiliation{Department of Physics, Indian Institute of Science, Bangalore 560 012, India.}
\begin{abstract}
Quantum dot lattices (QDLs) have the potential to allow for the tailoring of optical, magnetic and electronic properties of a user-defined artificial solid. We use a dual gated device structure to controllably tune the potential landscape in a GaAs/AlGaAs two-dimensional electron gas, thereby enabling the formation of a periodic QDL. The current-voltage characteristics, $I(V)$, follow a power law, as expected for a QDL. In addition, a systematic study of the scaling behavior of $I(V)$ allows us to probe the effects of background disorder on transport through the QDL. Our results are particularly important for semiconductor-based QDL architectures which aim to probe collective phenomena.

\end{abstract}


\maketitle

\section{Introduction}
A single quantum dot has a distinct energy spectrum and has often been termed an `artificial atom'. One may thus expect a periodic array of quantum dots to exhibit collective properties of an `artificial solid'. This has been a strong motivation for a wide variety of experimental and theoretical work on quantum dot lattices (QDLs). Two-dimensional (2D) QDLs have primarily been realized using self-assembled nanocrystal arrays.~\cite{Parthasarathy_PRL_2001, Parthasarathy_PRL_2004, Romero_PbSe_PRL_2005, Moreira_Cotunnuling_PRL_2011} Electrical transport in these systems is governed not only by electron-electron interactions but also by disorder. This is clearly reflected in transport studies, which go well beyond the observation of standard Coulomb blockade. For example, Parthasarathy \emph{et al.} have emphasized the important role that structural and charge disorder can play on the current-voltage [$I(V)$] characteristics of the system.~\cite{Parthasarathy_PRL_2001} It has also been shown recently that the transport through such arrays can exhibit cotunneling effects, leading to an Efros-Shklovskii type variable range hopping,~\cite{Moreira_Cotunnuling_PRL_2011} signifying the importance of Coulomb interactions in the system. Other routes toward fabricating 2D QDLs include lithographically defined metal islands~\cite{Rimberg_PRL, Kurdak_PRB_1998} and semiconductor quantum dot arrays.~\cite{Smith_JVAC_1992,Lorke_PRB_FIR_1992,Meurer_PRL_QuantumDotAtom_1992,Duruoz_PRL, Dorn_PRB}

With regard to the study of many-body physics, we expect semiconductor QDLs to exhibit a significant advantage due to primarily two reasons; (1) the inter-dot coupling and dot charging energy can be controlled \emph{in situ} using gate voltages, (2) the high Fermi temperatures in these systems should allow for the observation of Mott-Hubbard-like physics at experimentally accessible temperatures.~\cite{Byrnes_PRB} Early measurements using far-infrared spectroscopy were used to study the energy levels of individual quantum dots in a gate-defined QDL.~\cite{Meurer_PRL_QuantumDotAtom_1992} A similar method was employed to study the transition from an antidot lattice (ADL) to a QDL using a single grid-like gate.~\cite{Lorke_PRB_FIR_1992} Very recently, optical studies on a large array of etched quantum dots in a GaAs/AlGaAs two-dimensional electron gas (2DEG) have even shown the existence of Mott-Hubbard electrons.~\cite{Singha_Science} However, electrical transport through such semiconductor QDLs remains relatively unexplored. Smith \emph{et al.} used a patterned gate~\cite{Smith_JVAC_1992} to introduce a periodic potential in the 2DEG. Their magnetic field studies showed a splitting of Landau levels when the periodic potential was weak, and possible signatures of a QDL for stronger potentials. A similar device structure was used to study the combined effects of a weak periodic potential and strong disorder in large quantizing magnetic fields.~\cite{Melinte_PRL_2004} Duru\"oz \emph{et al.} were the first to study non-equilibrium transport through an \emph{etched} QDL with a significantly stronger confining potential.~\cite{Duruoz_PRL} They found $I(V)$ to follow a power law, in agreement with the predictions of Middleton and Wingreen (MW).~\cite{MW_PRL} However, they focused primarily on hysteresis and switching effects in the system. More recently, Dorn \emph{et al.} studied a similar system with quantum dots defined by a combination of atomic force microscope oxidation and a top gate.~\cite{Dorn_PRB} They did not appear to observe the standard power law dependence expected from a collective behavior of the QDL, but used percolation analysis and magneto-transport to establish transport through a `network' of quantum dots.

In this work we focus on the \emph{dc} transport through a tunable QDL in a GaAs/AlGaAs 2DEG, which is defined solely by electrostatic gating. The electrostatic definition of the QDL allows us to controllably alter the potential landscape in the 2DEG. It also avoids the inevitable introduction of additional disorder associated with etching and oxidation techniques. The versatility of the device structure enables us to drive the system from a delocalized ADL to a localized QDL by simply changing the gate voltage configuration. Furthermore, it allows us to decouple the effects of strong background disorder from transport through the QDL of interest, something which has not been possible thus far.
\begin{figure*}[!tbh]
\includegraphics[width=1\linewidth]{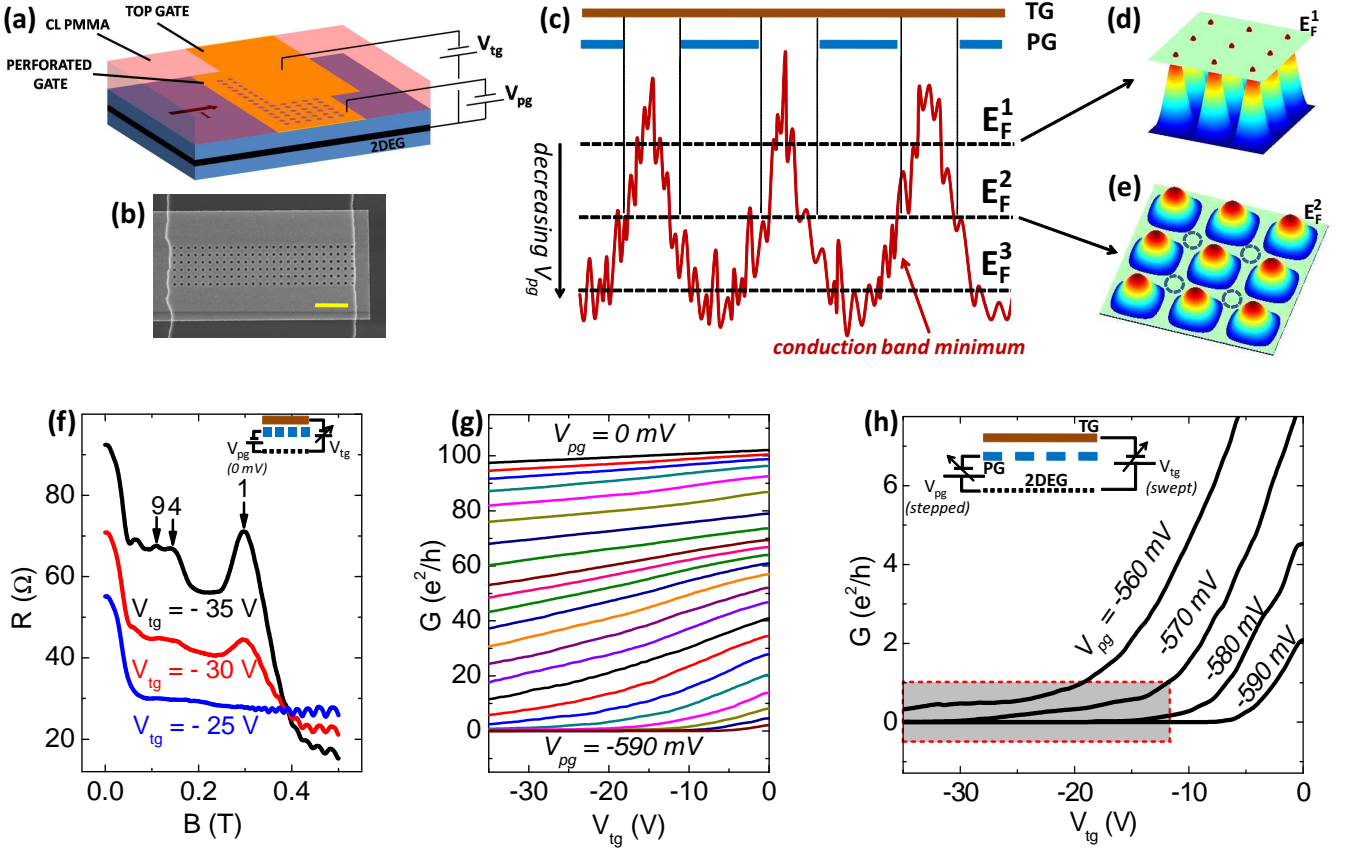}
\caption{(a) The dual gated device structure with a perforated gate (PG), followed by cross-linked polymethyl methacrylate (CL PMMA), and an un-patterned top gate (TG). Voltages on the TG ($V_{tg}$) and PG ($V_{pg}$) are used to control the potential landscape in the two-dimensional electron gas (2DEG). $I$ represents the direction of current flow. (b) Scanning electron micrograph of the perforated gate. The scale bar is $1$~$\mu$m. (c) Sketch of the potential profile below the gated region. The TG acts through the holes in the PG and introduces a strong periodic potential. In addition to this, there exist disorder induced fluctuations in the conduction band minimum. At a relatively high Fermi energy ($E_F$), indicated by $E_F^1$, an antidot lattice is formed (as suggested by the potential landscape in (d)). A reduction in $E_F$ to $E_F^2$ localizes the system, thus forming a quantum dot lattice (QDL), with a quantum dot formed between four adjacent antidots, as shown in (e). Finally, at $E_F^3$, the effects of disorder become significant, thus masking the effects of the QDL. (f) As $V_{tg}$ is reduced from -25~V to -35~V (with $V_{pg}=0$~mV), clear commensurability peaks in magneto-resistance become apparent, signifying the formation of an antidot lattice in a device similar to the one used to from the QDL. (g) Variation of equilibrium conductance ($G$) with top $V_{tg}$ for different values of $V_{pg}$. The top (bottom) trace corresponds to $V_{pg}=0$~mV ($-590$~mV). (h) A zoom-in on the bottom four traces from (g). The shaded region shows the section of the $V_{tg}$-$V_{pg}$ phase space where the QDL is realized.}
\label{fig1}
\end{figure*}
\section{Device Structure}
The 2DEG used for these studies was located 60~nm below the surface and the Si $\delta$-doping layer was located 30~nm above the 2DEG. The wafers had an as-grown sheet density ($n_s$) of $4\times 10^{15}$~m$^{-2}$, and low temperature mobility of $230$~m$^{2}$/Vs. The device structure is outlined in Figure~\ref{fig1}a. The first layer consists of a Ti/Au (20~nm/15~nm) perforated gate (PG) on the substrate. Figure~\ref{fig1}b shows a scanning electron micrograph of a typical PG. This is followed by a 150~nm thick cross-linked polymethyl methacrylate (CL PMMA) layer, which is obtained by exposing the desired portion of PMMA to a large electron beam dose. The CL PMMA serves as a dielectric layer and is resistant to further processing of the device. Finally, an un-patterned Ti/Au (25~nm/80~nm) top gate (TG) is deposited such that it lies directly above the PG. We present results on two kinds of devices. The first (dev1) has a PG which consists of a $6\times 26$ array of perforations with diameter, $d=150$~nm and center to center distance, $a=300$~nm.  The second (dev2) has $d=300$~nm and $a=700$~nm and an array of $3\times 11$ perforations. All measurements were performed in a He-3 cryostat with a base temperature of $300$~mK.

\begin{figure*}[!tbh]
\includegraphics[width=.7\linewidth]{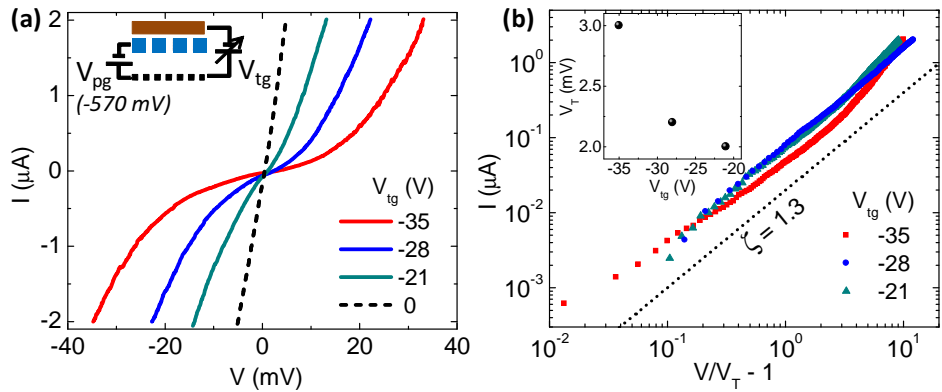}
\caption{(a) Introduction of non-linearity in $I(V)$ with a reduction in $V_{tg}$ from $-21$~V to $-35$~V, with $V_{pg}$ fixed at $-570$~mV. The dashed line shows that $I(V)$ is linear when $V_{tg}=0$. (b) $I$ vs.$V/V_T-1$ on a log-log plot showing a power law scaling for the curves in (a). The dotted line indicates $\zeta=1.3$. The inset shows the variation of the threshold voltage ($V_T$) with $V_{tg}$.}
\label{fig2}
\end{figure*}
\section{Results and Discussion}
By applying a large negative voltage on the TG ($V_{tg}$), regions in the 2DEG directly below the PG perforations are depleted, resulting in a periodic antidot potential which intersects the Fermi surface. This is shown schematically in Figure~\ref{fig1}c, where in addition to the periodic potential, there exist random fluctuations in the conduction band minimum that arise from background disorder. As long as $E_F$ is relatively high (depicted by $E_F^1$), the system is delocalized as a whole, and an antidot lattice is formed. Figure~\ref{fig1}d shows a three-dimensional schematic of the potential landscape in such a case. The application of a magnetic field now results in the emergence of commensurability peaks in resistance ($R$) arising from pinned orbits around the antidots~.\cite{Weiss_PRL_Commens_1991} Figure~\ref{fig1}f clearly shows that as $V_{tg}$ is lowered, the resistance maxima (associated with orbits around 1, 4, and 9 antidots) become more pronounced. The position of the fundamental commensurability peak (orbits around 1 antidot) may be expressed as $B_1=\frac{2\hbar\sqrt{2\pi n_s}}{ea}$, where $e$ is the electronic charge. Evaluating this for dev2, gives $B_1=0.32$~T, in very good agreement with the peak position in Figure~\ref{fig1}f. We have studied such antidot lattices in detail previously,~\cite{Goswami_APL} and have demonstrated a fine control over the potential landscape in the 2DEG. As $E_F$ is lowered to $E_F^2$ (Figure~\ref{fig1}c), the electrons become confined to the potential minima arising from the antidots, thereby forming a QDL, as depicted in Figure~\ref{fig1}e. A measurement of the equilibrium conductance ($G$) shows signatures of such a transition for dev1. We note that dev1 consists of smaller, more closely spaced perforations, which should make Coulomb blockade related effects more observable. We therefore focus on dev1 for the remainder of this work. Figure~\ref{fig1}g shows the behavior of $G$ (measured using standard low frequency lock-in techniques) as a function of $V_{tg}$ for different voltages on the PG ($V_{pg}$). As $V_{pg}$ is reduced ($E_F$ is reduced) from $0$~mV (top trace) to $-590$~mV (bottom trace), the action of the TG becomes significantly stronger. This suggests that when both $V_{tg}$ and $V_{pg}$ are sufficiently negative, a QDL is realized, where the quantum dots are formed between four adjacent antidots and are coupled to each other through the high inter-dot tunnel resistances of the resultant quasi one-dimensional channels (see Figure~\ref{fig1}e). The shaded region in Figure~\ref{fig1}h shows the portion of the $V_{tg}$-$V_{pg}$ phase space where such a situation is realized. Note that in this region we always have $G<e^2/h$, suggesting that the tunnel resistance between dots is larger than the resistance quantum.

\begin{figure*}[!tbh]
\includegraphics[width=1\linewidth]{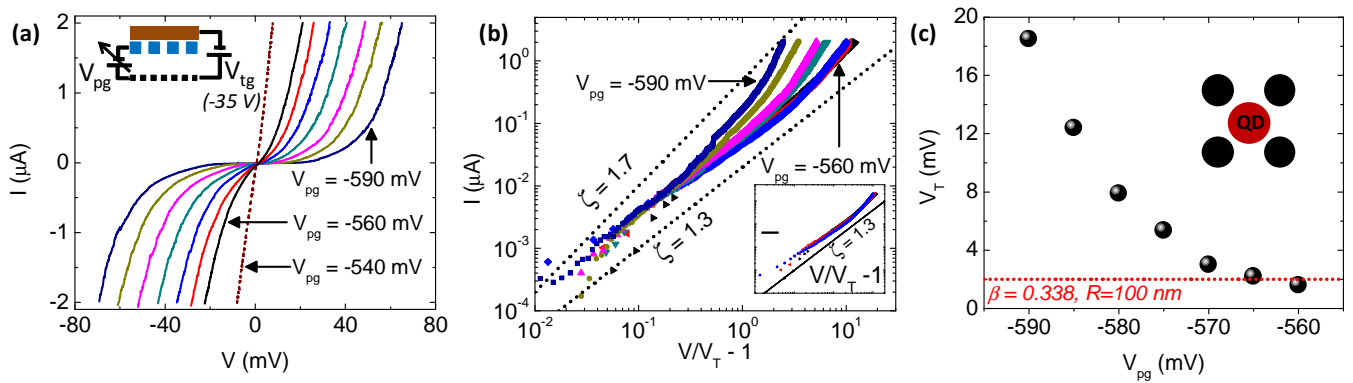}
\caption{(a) $I(V)$ for $V_{pg}$ ranging from $-560$~mV to $-590$~mV in steps of $5$~mV, with $V_{tg}$ fixed at $-35$~V. The dashed line indicates that $I(V)$ is linear when $V_{pg}=-540$~mV. (b) $I(V/V_T-1)$ for the curves in (a) showing a power law. The inset shows the collapse of these curves at higher $V_{pg}$. The dotted lines indicate an increase in $\zeta$ (from $1.3$ to $1.7$) with a reduction in $V_{pg}$. (c) Variation of the threshold voltage ($V_T$) with $V_{pg}$. Dotted line shows a rough estimate of the expected value of $V_T$ for our system. Inset shows a schematic (to scale) of the relative size of the quantum dot (QD) and its neighboring antidots, based on the lithographic dimensions of the device.}
\label{fig3}
\end{figure*}
A similar transition from a delocalized state to a localized one has been reported in semiconductor quantum dot arrays.~\cite{Dorn_PRB, Simoni_APL} However, neither of these studies concentrated on the $I(V)$ characteristics of the system. To validate our description of the formation of a QDL outlined above, we perform a systematic study of the $I(V)$ characteristics of the device. We apply a symmetric bias across the device ($+V/2$ at one end and $-V/2$ at the other), and the current is measured using a Keithley 2400 source measure unit. When the device is tuned to the QDL regime, we expect $I(V)$ to obey the MW model~\cite{MW_PRL} and
\begin{equation}
\label{eq1}
I\sim (V/V_T-1)^{\zeta}\:,
\end{equation}
where $V_T$ is the threshold voltage below which the current is suppressed, and $\zeta$ is the power law exponent. Figure~\ref{fig2}a shows that the device is perfectly ohmic when $V_{pg}=-570$~mV (indicating a relatively high $E_F$) and $V_{tg}=0$~V (dashed line). As $V_{tg}$ is reduced to $-21$~V, there is an observable deviation from linearity, characteristic of Coulomb blockade. A further reduction in $V_{tg}$ strengthens this non-linearity. In Figure~\ref{fig2}b we plot $I$ vs. $V/V_T-1$ (the normalized voltage). According to Equation~\ref{eq1}, we expect multiple curves with different values of $V_T$ to collapse onto a single trace with a unique $\zeta$. Our curves show a reasonable collapse, with a clear power law characterized by $\zeta\sim 1.3$. The absence of a sharp threshold in Figure~\ref{fig2}a makes it difficult to directly identify $V_T$. However, the shape of the curves in Figure~\ref{fig2}b are highly sensitive to the choice of $V_T$. We therefore adopt the procedure followed by Rimberg \emph{et al.}~\cite{Rimberg_PRL} and choose $V_T$ such that the traces exhibit a power law over the largest possible range. This results in a relatively small ($<3\%$) error in the estimation of $V_T$. We note that at lower values of $V_{pg}$, the threshold becomes more apparent and it is possible to easily identify $V_T$. However, as we will show later, the effects of background disorder also become more pronounced at lower $V_{pg}$, and it is not possible to obtain the collapse observed in Figure~\ref{fig2}b. At any rate, we can certainly conclude that the introduction of the periodic potential using $V_{tg}$ results in $I(V)$ that exhibits a clear power law, characteristic of a QDL. We note that sweeping the bias voltage up and down did not result in any hysteresis, as observed in some previous work.~\cite{Duruoz_PRL} Also, there were no observable leakage currents ($<1$~pA) for all the gate voltage configurations used in this study.

If our picture of the potential landscape is indeed correct, we should also be able to induce a QDL by keeping $V_{tg}$ fixed, and lowering $V_{pg}$. Figure~\ref{fig3}a shows precisely this. The dashed line indicates that the $I(V)$ is linear when $E_F$ is high ($V_{pg}=-540$~mV), with $V_{tg}=-35$~V. As $V_{pg}$ is reduced from $-560$~mV to $-590$~mV, the non-linearity becomes more apparent, and we see the emergence of a clear $V_T$. Figure~\ref{fig3}b shows that all the curves exhibit a clear power law, however a collapse according to Equation~\ref{eq1} is only possible for higher values of $V_{pg}$ (shown separately in the inset for clarity). Again, we find $\zeta\sim 1.3$, as observed in Figure~\ref{fig3}b, reinforcing the fact that both sets of gate voltage configurations yield a QDL with similar characteristics. We also observe a slight upturn at higher values of $I$, but are not certain about its origin at present. Below $V_{pg}=-570$~mV, $\zeta$ steadily increases to $\sim 1.7$. In addition, $V_T$ increases quite rapidly from $3.0$~mV to $18.5$~mV as $V_{pg}$ is reduced from $-570$~mV to $-590$~mV (Figure~\ref{fig3}c).

It has been shown that an increase in background charge disorder can significantly enhance $V_T$.~\cite{Muller_JAP_Disorder_1998} In a 2DEG, such disorder arises primarily due to potential fluctuations from the dopant layer. As $E_F$ is reduced, the ability of the 2DEG to screen the disorder becomes progressively weaker, thereby increasing the amount of charge variation within the QDL. We believe that this accounts for the observation of the rapidly increasing $V_T$. We expect $V_T\approx NE_C\beta $, where $N$ is the number of quantum dots in the direction of transport, $E_C$ is the charging energy of each dot, and $\beta$ is a factor which depends on the lattice geometry and the relevant capacitances in the system.~\cite{MW_PRL} In the limit of weak inter-dot coupling, $\beta=0.338$ for a square lattice. In the absence of a precise knowledge of the potential landscape, we use the lithographic dimensions as a guide to estimate $V_T\sim 2$~mV [dashed line in Figure~\ref{fig3}c], assuming a quantum dot radius of $100$~nm. This is sketched as a (to scale) schematic in the inset of Figure~\ref{fig3}c. Note that $V_T$ increases by almost $10$ times as $V_{pg}$ is reduced from $-560$~mV to $-590$~mV. Assuming that all other parameters remain constant, this would imply that the dot radius reduces by a factor of 10 and approaches $\sim 10$~nm, which seems unlikely. Thus, the behavior of $V_T$ cannot be explained by just a variation in the properties of the quantum dots and their coupling. This supports the scenario where, at significantly low values of $E_F$, transport through the previously uniform QDL has an added contribution from disorder. This effectively increases the energy required for an electron to move from one end to the other, thus increasing $V_T$.
\begin{figure*}[!tbh]
\includegraphics[width=1\linewidth]{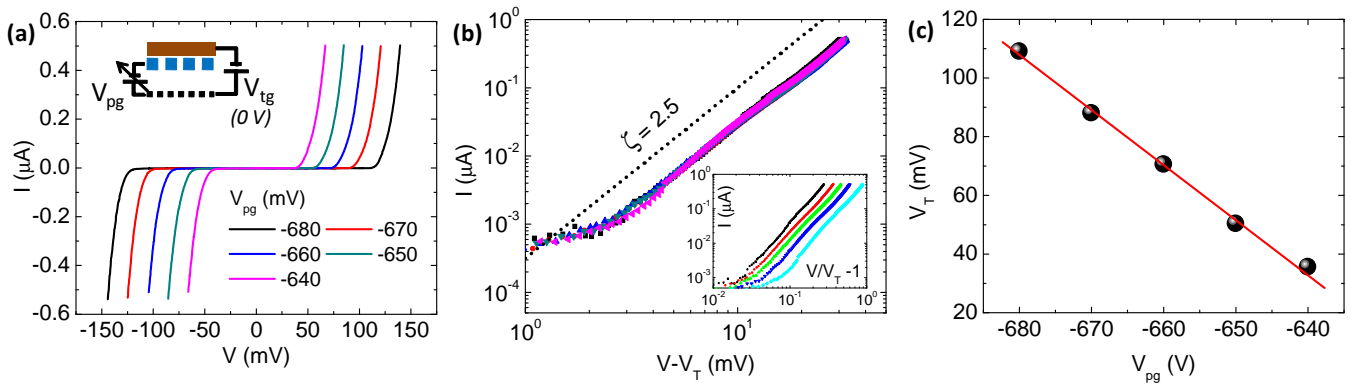}
\caption{(a) $I(V)$ at very low values of $V_{pg}$ ($\leq 640$~mV) displaying a large blockaded region and a distinct threshold. (b) Collapse of curves in (a) obtained by re-scaling the voltage axis to $(V-V_T)$. All curves display $\zeta\sim 2.5$. The inset shows the failure of a collapse for $I(V/V_T-1)$ traces. (c) Linear dependence of $V_T$ on $V_{pg}$.}
\label{fig4}
\end{figure*}

To test this hypothesis, we set $V_{tg}=0$ and take $V_{pg}$ to large negative values, where the zero bias conductance is zero. Under these circumstances, when we have driven the system well below the percolation threshold, there should be no significant contribution from the QDL. Instead, we expect the transport to be completely dominated by an array of disconnected electron puddles with a distribution of sizes.~\cite{DasSarma_Percolation_PRL_2005} Figure~\ref{fig4}a shows $I(V)$ for different values of $V_{pg}$. Not only do these curves have a very well-defined $V_T$, but all of them can be characterized by a single $\zeta\sim 2.5$ (Figure~\ref{fig4}b). This indicates that the observed increase in $\zeta$ and $V_T$ in Figure~\ref{fig3}b and c respectively, is directly related to the increasing effect of disorder. From the inset of Figure~\ref{fig4}b it is clear that in this transport regime it is impossible to collapse the traces using the standard MW model described by Equation~\ref{eq1}. However, a universal curve is obtained if each trace is translated by an amount $-V_T$ (Figure~\ref{fig4}b). Furthermore, Figure~\ref{fig4}c shows that $V_T$ varies linearly with $V_{pg}$, with much larger values of $V_T$. It is worth noting that these results are strikingly similar to those of Parthasarathy \emph{et al.},~\cite{Parthasarathy_PRL_2004} the only difference being that they observed a similar dependence of the $I(V)$ on \emph{temperature} ($T$), rather than gate voltage. They ascribed the linear dependence of $V_T$ on $T$ to the existence of a distribution of charging energies in a highly disordered QDL. As the temperature increases, certain quantum dots become transparent and do not contribute to $V_T$. A detailed analysis based on results from percolation theory was able to explain the linear relationship between $V_T$ and $T$, and a collapse of $I$ vs $(V-V_T)$ traces. A similar distribution of charging energies is quite probable in a highly inhomogeneous 2DEG. When $E_F$ is increased, certain inter-dot barriers are effectively overcome, resulting in a reduction in $V_T$. We expect this to continue up to some characteristic $E_F$, above which the system looks fairly homogeneous. Though this description is qualitatively in line with our observations, more detailed calculations are required to verify the validity of the model. An alternative explanation for the linearity in $V_T$ may lie in the fact that at extremely low values of $V_{pg}$, the Fermi level lies well below the conduction band minimum. The application of a large bias voltage then induces carriers, resulting in the onset of conduction at $V_T$. In this scenario, one would also expect $V_T$ to be proportional to $V_{pg}$. Further experiments are required to understand this scaling of the $I(V)$ characteristics at significantly lower values of $V_{pg}$.
\\
\section{Conclusions}
In conclusion, we have studied non-equilibrium transport through an electrostatically defined quantum dot lattice. Our dual gated device structure has a significant advantage over previous designs in that it allows us to decouple the effects of strong background disorder from those arising solely due to the QDL. These effects are reflected in an increase in the threshold voltage as well as a deviation from the standard MW scaling. Our results are particularly important for semiconductor-based QDL architectures which aim to probe collective phenomena using transport.

\section*{ACKNOWLEDGEMENTS}
We would like to acknowledge support from UKIERI and EPSRC. I.F. acknowledges Toshiba Research Europe for financial support. S.G. thanks the IISc Centenary Postdoctoral Fellowship. We are grateful to Nigel Cooper and Daniel Thompson for useful discussions.


\begin{thebibliography}{20}
\expandafter\ifx\csname natexlab\endcsname\relax\def\natexlab#1{#1}\fi
\expandafter\ifx\csname bibnamefont\endcsname\relax
  \def\bibnamefont#1{#1}\fi
\expandafter\ifx\csname bibfnamefont\endcsname\relax
  \def\bibfnamefont#1{#1}\fi
\expandafter\ifx\csname citenamefont\endcsname\relax
  \def\citenamefont#1{#1}\fi
\expandafter\ifx\csname url\endcsname\relax
  \def\url#1{\texttt{#1}}\fi
\expandafter\ifx\csname urlprefix\endcsname\relax\def\urlprefix{URL }\fi
\providecommand{\bibinfo}[2]{#2}
\providecommand{\eprint}[2][]{\url{#2}}

\bibitem[{\citenamefont{Parthasarathy et~al.}(2001)\citenamefont{Parthasarathy,
  Lin, and Jaeger}}]{Parthasarathy_PRL_2001}
\bibinfo{author}{\bibfnamefont{R.}~\bibnamefont{Parthasarathy}},
  \bibinfo{author}{\bibfnamefont{X.-M.} \bibnamefont{Lin}}, \bibnamefont{and}
  \bibinfo{author}{\bibfnamefont{H.~M.} \bibnamefont{Jaeger}},
  \bibinfo{journal}{Phys. Rev. Lett.} \textbf{\bibinfo{volume}{87}},
  \bibinfo{pages}{186807} (\bibinfo{year}{2001}).

\bibitem[{\citenamefont{Parthasarathy et~al.}(2004)\citenamefont{Parthasarathy,
  Lin, Elteto, Rosenbaum, and Jaeger}}]{Parthasarathy_PRL_2004}
\bibinfo{author}{\bibfnamefont{R.}~\bibnamefont{Parthasarathy}},
  \bibinfo{author}{\bibfnamefont{X.-M.} \bibnamefont{Lin}},
  \bibinfo{author}{\bibfnamefont{K.}~\bibnamefont{Elteto}},
  \bibinfo{author}{\bibfnamefont{T.~F.} \bibnamefont{Rosenbaum}},
  \bibnamefont{and} \bibinfo{author}{\bibfnamefont{H.~M.}
  \bibnamefont{Jaeger}}, \bibinfo{journal}{Phys. Rev. Lett.}
  \textbf{\bibinfo{volume}{92}}, \bibinfo{pages}{076801}
  (\bibinfo{year}{2004}).

\bibitem[{\citenamefont{Romero and Drndic}(2005)}]{Romero_PbSe_PRL_2005}
\bibinfo{author}{\bibfnamefont{H.~E.} \bibnamefont{Romero}} \bibnamefont{and}
  \bibinfo{author}{\bibfnamefont{M.}~\bibnamefont{Drndic}},
  \bibinfo{journal}{Phys. Rev. Lett.} \textbf{\bibinfo{volume}{95}},
  \bibinfo{pages}{156801} (\bibinfo{year}{2005}).

\bibitem[{\citenamefont{Moreira et~al.}(2011)\citenamefont{Moreira, Yu, Nadal,
  Bresson, Rosticher, Lequeux, Zimmers, and
  Aubin}}]{Moreira_Cotunnuling_PRL_2011}
\bibinfo{author}{\bibfnamefont{H.}~\bibnamefont{Moreira}},
  \bibinfo{author}{\bibfnamefont{Q.}~\bibnamefont{Yu}},
  \bibinfo{author}{\bibfnamefont{B.}~\bibnamefont{Nadal}},
  \bibinfo{author}{\bibfnamefont{B.}~\bibnamefont{Bresson}},
  \bibinfo{author}{\bibfnamefont{M.}~\bibnamefont{Rosticher}},
  \bibinfo{author}{\bibfnamefont{N.}~\bibnamefont{Lequeux}},
  \bibinfo{author}{\bibfnamefont{A.}~\bibnamefont{Zimmers}}, \bibnamefont{and}
  \bibinfo{author}{\bibfnamefont{H.}~\bibnamefont{Aubin}},
  \bibinfo{journal}{Phys. Rev. Lett.} \textbf{\bibinfo{volume}{107}},
  \bibinfo{pages}{176803} (\bibinfo{year}{2011}).

\bibitem[{\citenamefont{Rimberg et~al.}(1995)\citenamefont{Rimberg, Ho, and
  Clarke}}]{Rimberg_PRL}
\bibinfo{author}{\bibfnamefont{A.~J.} \bibnamefont{Rimberg}},
  \bibinfo{author}{\bibfnamefont{T.~R.} \bibnamefont{Ho}}, \bibnamefont{and}
  \bibinfo{author}{\bibfnamefont{J.}~\bibnamefont{Clarke}},
  \bibinfo{journal}{Phys. Rev. Lett.} \textbf{\bibinfo{volume}{74}},
  \bibinfo{pages}{4714} (\bibinfo{year}{1995}).

\bibitem[{\citenamefont{Kurdak et~al.}(1998)\citenamefont{Kurdak, Rimberg, Ho,
  and Clarke}}]{Kurdak_PRB_1998}
\bibinfo{author}{\bibfnamefont{C.}~\bibnamefont{Kurdak}},
  \bibinfo{author}{\bibfnamefont{A.~J.} \bibnamefont{Rimberg}},
  \bibinfo{author}{\bibfnamefont{T.~R.} \bibnamefont{Ho}}, \bibnamefont{and}
  \bibinfo{author}{\bibfnamefont{J.}~\bibnamefont{Clarke}},
  \bibinfo{journal}{Phys. Rev. B} \textbf{\bibinfo{volume}{57}},
  \bibinfo{pages}{R6842} (\bibinfo{year}{1998}).

\bibitem[{\citenamefont{Smith et~al.}(1992)\citenamefont{Smith, Chen, Pepper,
  Ahmed, Hasko, Ritchie, Frost, and Jones}}]{Smith_JVAC_1992}
\bibinfo{author}{\bibfnamefont{C.~G.} \bibnamefont{Smith}},
  \bibinfo{author}{\bibfnamefont{W.}~\bibnamefont{Chen}},
  \bibinfo{author}{\bibfnamefont{M.}~\bibnamefont{Pepper}},
  \bibinfo{author}{\bibfnamefont{H.}~\bibnamefont{Ahmed}},
  \bibinfo{author}{\bibfnamefont{D.}~\bibnamefont{Hasko}},
  \bibinfo{author}{\bibfnamefont{D.~A.} \bibnamefont{Ritchie}},
  \bibinfo{author}{\bibfnamefont{J.~E.~F.} \bibnamefont{Frost}},
  \bibnamefont{and} \bibinfo{author}{\bibfnamefont{G.~A.~C.}
  \bibnamefont{Jones}}, \bibinfo{journal}{J. Vac. Sci. Technol. B}
  \textbf{\bibinfo{volume}{10}}, \bibinfo{pages}{2904} (\bibinfo{year}{1992}).

\bibitem[{\citenamefont{Lorke et~al.}(1992)\citenamefont{Lorke, Jejina, and
  Kotthaus}}]{Lorke_PRB_FIR_1992}
\bibinfo{author}{\bibfnamefont{A.}~\bibnamefont{Lorke}},
  \bibinfo{author}{\bibfnamefont{I.}~\bibnamefont{Jejina}}, \bibnamefont{and}
  \bibinfo{author}{\bibfnamefont{J.~P.} \bibnamefont{Kotthaus}},
  \bibinfo{journal}{Phys. Rev. B} \textbf{\bibinfo{volume}{46}},
  \bibinfo{pages}{12845} (\bibinfo{year}{1992}).

\bibitem[{\citenamefont{Meurer et~al.}(1992)\citenamefont{Meurer, Heitmann, and
  Ploog}}]{Meurer_PRL_QuantumDotAtom_1992}
\bibinfo{author}{\bibfnamefont{B.}~\bibnamefont{Meurer}},
  \bibinfo{author}{\bibfnamefont{D.}~\bibnamefont{Heitmann}}, \bibnamefont{and}
  \bibinfo{author}{\bibfnamefont{K.}~\bibnamefont{Ploog}},
  \bibinfo{journal}{Phys. Rev. Lett.} \textbf{\bibinfo{volume}{68}},
  \bibinfo{pages}{1371} (\bibinfo{year}{1992}).

\bibitem[{\citenamefont{Duru\"oz et~al.}(1995)\citenamefont{Duru\"oz, Clarke,
  Marcus, and Harris}}]{Duruoz_PRL}
\bibinfo{author}{\bibfnamefont{C.~I.} \bibnamefont{Duru\"oz}},
  \bibinfo{author}{\bibfnamefont{R.~M.} \bibnamefont{Clarke}},
  \bibinfo{author}{\bibfnamefont{C.~M.} \bibnamefont{Marcus}},
  \bibnamefont{and} \bibinfo{author}{\bibfnamefont{J.~S.} \bibnamefont{Harris},
  \bibfnamefont{Jr.}}, \bibinfo{journal}{Phys. Rev. Lett.}
  \textbf{\bibinfo{volume}{74}}, \bibinfo{pages}{3237} (\bibinfo{year}{1995}).

\bibitem[{\citenamefont{Dorn et~al.}(2004)\citenamefont{Dorn, Ihn, Ensslin,
  Wegscheider, and Bichler}}]{Dorn_PRB}
\bibinfo{author}{\bibfnamefont{A.}~\bibnamefont{Dorn}},
  \bibinfo{author}{\bibfnamefont{T.}~\bibnamefont{Ihn}},
  \bibinfo{author}{\bibfnamefont{K.}~\bibnamefont{Ensslin}},
  \bibinfo{author}{\bibfnamefont{W.}~\bibnamefont{Wegscheider}},
  \bibnamefont{and} \bibinfo{author}{\bibfnamefont{M.}~\bibnamefont{Bichler}},
  \bibinfo{journal}{Phys. Rev. B} \textbf{\bibinfo{volume}{70}},
  \bibinfo{pages}{205306} (\bibinfo{year}{2004}).

\bibitem[{\citenamefont{Byrnes et~al.}(2008)\citenamefont{Byrnes, Kim, Kusudo,
  and Yamamoto}}]{Byrnes_PRB}
\bibinfo{author}{\bibfnamefont{T.}~\bibnamefont{Byrnes}},
  \bibinfo{author}{\bibfnamefont{N.~Y.} \bibnamefont{Kim}},
  \bibinfo{author}{\bibfnamefont{K.}~\bibnamefont{Kusudo}}, \bibnamefont{and}
  \bibinfo{author}{\bibfnamefont{Y.}~\bibnamefont{Yamamoto}},
  \bibinfo{journal}{Phys. Rev. B} \textbf{\bibinfo{volume}{78}},
  \bibinfo{pages}{075320} (\bibinfo{year}{2008}).

\bibitem[{\citenamefont{Singha et~al.}(2011)\citenamefont{Singha, Gibertini,
  Karmakar, Yuan, Polini, Vignale, Katsnelson, Pinczuk, Pfeiffer, West
  et~al.}}]{Singha_Science}
\bibinfo{author}{\bibfnamefont{A.}~\bibnamefont{Singha}},
  \bibinfo{author}{\bibfnamefont{M.}~\bibnamefont{Gibertini}},
  \bibinfo{author}{\bibfnamefont{B.}~\bibnamefont{Karmakar}},
  \bibinfo{author}{\bibfnamefont{S.}~\bibnamefont{Yuan}},
  \bibinfo{author}{\bibfnamefont{M.}~\bibnamefont{Polini}},
  \bibinfo{author}{\bibfnamefont{G.}~\bibnamefont{Vignale}},
  \bibinfo{author}{\bibfnamefont{M.~I.} \bibnamefont{Katsnelson}},
  \bibinfo{author}{\bibfnamefont{A.}~\bibnamefont{Pinczuk}},
  \bibinfo{author}{\bibfnamefont{L.~N.} \bibnamefont{Pfeiffer}},
  \bibinfo{author}{\bibfnamefont{K.~W.} \bibnamefont{West}},
  \bibnamefont{\emph{et~al.}}, \bibinfo{journal}{Science}
  \textbf{\bibinfo{volume}{332}}, \bibinfo{pages}{1176} (\bibinfo{year}{2011}).

\bibitem[{\citenamefont{Melinte et~al.}(2004)\citenamefont{Melinte, Berciu,
  Zhou, Tutuc, Papadakis, Harrison, De~Poortere, Wu, Chaikin, Shayegan
  et~al.}}]{Melinte_PRL_2004}
\bibinfo{author}{\bibfnamefont{S.}~\bibnamefont{Melinte}},
  \bibinfo{author}{\bibfnamefont{M.}~\bibnamefont{Berciu}},
  \bibinfo{author}{\bibfnamefont{C.}~\bibnamefont{Zhou}},
  \bibinfo{author}{\bibfnamefont{E.}~\bibnamefont{Tutuc}},
  \bibinfo{author}{\bibfnamefont{S.~J.} \bibnamefont{Papadakis}},
  \bibinfo{author}{\bibfnamefont{C.}~\bibnamefont{Harrison}},
  \bibinfo{author}{\bibfnamefont{E.~P.} \bibnamefont{De~Poortere}},
  \bibinfo{author}{\bibfnamefont{M.}~\bibnamefont{Wu}},
  \bibinfo{author}{\bibfnamefont{P.~M.} \bibnamefont{Chaikin}},
  \bibinfo{author}{\bibfnamefont{M.}~\bibnamefont{Shayegan}},
  \bibnamefont{\emph{et~al.}}, \bibinfo{journal}{Phys. Rev. Lett.}
  \textbf{\bibinfo{volume}{92}}, \bibinfo{pages}{036802}
  (\bibinfo{year}{2004}).

\bibitem[{\citenamefont{Middleton and Wingreen}(1993)}]{MW_PRL}
\bibinfo{author}{\bibfnamefont{A.~A.} \bibnamefont{Middleton}}
  \bibnamefont{and} \bibinfo{author}{\bibfnamefont{N.~S.}
  \bibnamefont{Wingreen}}, \bibinfo{journal}{Phys. Rev. Lett.}
  \textbf{\bibinfo{volume}{71}}, \bibinfo{pages}{3198} (\bibinfo{year}{1993}).

\bibitem[{\citenamefont{Weiss et~al.}(1991)\citenamefont{Weiss, Roukes,
  Menschig, Grambow, von Klitzing, and Weimann}}]{Weiss_PRL_Commens_1991}
\bibinfo{author}{\bibfnamefont{D.}~\bibnamefont{Weiss}},
  \bibinfo{author}{\bibfnamefont{M.~L.} \bibnamefont{Roukes}},
  \bibinfo{author}{\bibfnamefont{A.}~\bibnamefont{Menschig}},
  \bibinfo{author}{\bibfnamefont{P.}~\bibnamefont{Grambow}},
  \bibinfo{author}{\bibfnamefont{K.}~\bibnamefont{von Klitzing}},
  \bibnamefont{and} \bibinfo{author}{\bibfnamefont{G.}~\bibnamefont{Weimann}},
  \bibinfo{journal}{Phys. Rev. Lett.} \textbf{\bibinfo{volume}{66}},
  \bibinfo{pages}{2790} (\bibinfo{year}{1991}).

\bibitem[{\citenamefont{Goswami et~al.}(2010)\citenamefont{Goswami, Siegert,
  Shamim, Pepper, Farrer, Ritchie, and Ghosh}}]{Goswami_APL}
\bibinfo{author}{\bibfnamefont{S.}~\bibnamefont{Goswami}},
  \bibinfo{author}{\bibfnamefont{C.}~\bibnamefont{Siegert}},
  \bibinfo{author}{\bibfnamefont{S.}~\bibnamefont{Shamim}},
  \bibinfo{author}{\bibfnamefont{M.}~\bibnamefont{Pepper}},
  \bibinfo{author}{\bibfnamefont{I.}~\bibnamefont{Farrer}},
  \bibinfo{author}{\bibfnamefont{D.~A.} \bibnamefont{Ritchie}},
  \bibnamefont{and} \bibinfo{author}{\bibfnamefont{A.}~\bibnamefont{Ghosh}},
  \bibinfo{journal}{Appl. Phys. Lett.} \textbf{\bibinfo{volume}{97}},
  \bibinfo{pages}{132104} (\bibinfo{year}{2010}).

\bibitem[{\citenamefont{Simoni et~al.}(2010)\citenamefont{Simoni, Singha,
  Gibertini, Karmakar, Polini, Piazza, Pfeiffer, West, Beltram, and
  Pellegrini}}]{Simoni_APL}
\bibinfo{author}{\bibfnamefont{G.~D.} \bibnamefont{Simoni}},
  \bibinfo{author}{\bibfnamefont{A.}~\bibnamefont{Singha}},
  \bibinfo{author}{\bibfnamefont{M.}~\bibnamefont{Gibertini}},
  \bibinfo{author}{\bibfnamefont{B.}~\bibnamefont{Karmakar}},
  \bibinfo{author}{\bibfnamefont{M.}~\bibnamefont{Polini}},
  \bibinfo{author}{\bibfnamefont{V.}~\bibnamefont{Piazza}},
  \bibinfo{author}{\bibfnamefont{L.~N.} \bibnamefont{Pfeiffer}},
  \bibinfo{author}{\bibfnamefont{K.~W.} \bibnamefont{West}},
  \bibinfo{author}{\bibfnamefont{F.}~\bibnamefont{Beltram}}, \bibnamefont{and}
  \bibinfo{author}{\bibfnamefont{V.}~\bibnamefont{Pellegrini}},
  \bibinfo{journal}{Appl. Phys. Lett.} \textbf{\bibinfo{volume}{97}},
  \bibinfo{pages}{132113} (\bibinfo{year}{2010}).

\bibitem[{\citenamefont{Muller et~al.}(1998)\citenamefont{Muller, Katayama, and
  Mizuta}}]{Muller_JAP_Disorder_1998}
\bibinfo{author}{\bibfnamefont{H.-O.} \bibnamefont{Muller}},
  \bibinfo{author}{\bibfnamefont{K.}~\bibnamefont{Katayama}}, \bibnamefont{and}
  \bibinfo{author}{\bibfnamefont{H.}~\bibnamefont{Mizuta}},
  \bibinfo{journal}{Journal of Applied Physics} \textbf{\bibinfo{volume}{84}},
  \bibinfo{pages}{5603 } (\bibinfo{year}{1998}).

\bibitem[{\citenamefont{Das~Sarma et~al.}(2005)\citenamefont{Das~Sarma, Lilly,
  Hwang, Pfeiffer, West, and Reno}}]{DasSarma_Percolation_PRL_2005}
\bibinfo{author}{\bibfnamefont{S.}~\bibnamefont{Das~Sarma}},
  \bibinfo{author}{\bibfnamefont{M.~P.} \bibnamefont{Lilly}},
  \bibinfo{author}{\bibfnamefont{E.~H.} \bibnamefont{Hwang}},
  \bibinfo{author}{\bibfnamefont{L.~N.} \bibnamefont{Pfeiffer}},
  \bibinfo{author}{\bibfnamefont{K.~W.} \bibnamefont{West}}, \bibnamefont{and}
  \bibinfo{author}{\bibfnamefont{J.~L.} \bibnamefont{Reno}},
  \bibinfo{journal}{Phys. Rev. Lett.} \textbf{\bibinfo{volume}{94}},
  \bibinfo{pages}{136401} (\bibinfo{year}{2005}).

\end{thebibliography}
\end{document}